\documentclass[iop, apjl]{emulateapj}
\usepackage[backref,breaklinks,colorlinks,citecolor=blue]{hyperref}
\usepackage{amsmath,amstext}
\usepackage[T1]{fontenc}
\usepackage{apjfonts}
\usepackage[figure,figure*]{hypcap}

\usepackage{graphicx}
\usepackage{color}
\DeclareGraphicsExtensions{.eps,.png,.jpg}
\usepackage{epsfig, psfig}
\usepackage[normalem]{ulem}

\begin{document}

\title{C\MakeLowercase{oordinated} A\MakeLowercase{ssembly of} 
G\MakeLowercase{alaxy} G\MakeLowercase{roups and} C\MakeLowercase{lusters 
in the} I\MakeLowercase{llustris}TNG S\MakeLowercase{imulations}}
\author{
  Meng Gu\altaffilmark{1}, 
  Charlie Conroy\altaffilmark{2}, 
  Benedikt Diemer\altaffilmark{3},
  Lars Hernquist\altaffilmark{2},
  Federico Marinacci\altaffilmark{4},
  Dylan Nelson\altaffilmark{5},
  R\"{u}diger Pakmor\altaffilmark{5},
  Annalisa Pillepich\altaffilmark{6},
  Mark Vogelsberger\altaffilmark{7}}  

\altaffiltext{1}{Department of Astrophysical Sciences, Princeton University, Princeton, NJ 08544, USA}
\altaffiltext{2}{Department of Astronomy, Harvard University, Cambridge, MA 02138, USA}
\altaffiltext{3}{Department of Astronomy, The University of Maryland, College Park, MD 20742}
\altaffiltext{4}{Department of Physics and Astronomy, University of Bologna, Via Gobetti 93/2, I-40129, Bologna, Italy}
\altaffiltext{5}{Max-Planck-Institut f\"{u}r Astrophysik, Karl-Schwarzschild-Str. 1, D-85748, Garching, Germany}
\altaffiltext{6}{Max-Planck-Institut f\"{u}r Astronomie, K\"{o}nigstuhl 17, D-69117 Heidelberg, Germany}
\altaffiltext{7}{Department of Physics, Massachusetts Institute of Technology, Cambridge, MA 02139}

\submitted{Submitted to ApJL}

\begin{abstract}

Recent stellar population analysis of early-type galaxy spectra has demonstrated 
that the low-mass galaxies in cluster centers have high [$\alpha/\rm Fe$] and 
old ages characteristic of massive galaxies and {\it unlike} the low-mass 
galaxy population in the outskirts of clusters and fields.  This phenomenon has 
been termed ``coordinated assembly'' to highlight the fact that the building 
blocks of massive cluster central galaxies are drawn from a special subset of 
the overall low-mass galaxy population.  Here we explore this idea in the 
IllustrisTNG simulations, particularly the TNG300 run, in order to understand 
how environment, especially cluster centers, shape the star formation histories 
of quiescent satellite galaxies in groups and clusters 
($M_{200c,z=0}\geq10^{13} M_{\odot}$).  Tracing histories of quenched satellite 
galaxies with $M_{\star,z=0}\geq10^{10} M_{\odot}$, we find that those in more 
massive dark matter halos, and located closer to the primary galaxies, are 
quenched earlier, have shorter star formation timescales, and older stellar 
ages.  The star formation timescale-$M_{\star}$ and stellar age-$M_{\star}$ 
scaling relations are in good agreement with observations, and are predicted 
to vary with halo mass and cluster-centric distance.  The dependence on 
environment arises due to the infall histories of satellite galaxies: 
galaxies that are located closer to cluster centers in more massive dark 
matter halos at $z=0$ were accreted earlier on average.  The delay between 
infall and quenching time is shorter for galaxies in more massive halos, and 
depends on the halo mass at its first accretion, showing that group 
pre-processing is a crucial aspect in satellite quenching.
\end{abstract}


\maketitle
\section{Introduction}
Recent observations and simulations provide us an increasingly clear picture of 
the build-up of massive early type galaxies (ETGs) 
\citep{Naab2009, Oser2010, Oser2012, vanDokkum2010, Patel2013}.  At high-$z$, 
strong dissipational processes such as gas accretion and gas rich mergers 
lead to rapid star formation, while at lower redshifts the accretion of low-mass systems 
dominate their growth.  The disrupted low-mass galaxies are predicted to leave 
their imprint, such as stellar population information, on massive 
galaxies, especially in their outer envelopes \citep{DiMatteo2009, Cook2016}.  
Meanwhile, observations of nearby ETGs reveal scaling relations between their 
stellar mass and stellar populations, e.g., stellar metallicity, 
stellar age, $\alpha$-abundance \citep{Trager2000, Thomas2005, Conroy2014}.  
Whether the building blocks of massive galaxies are intrinsically the same as 
the low-mass galaxies we observe today is still in debate.

\begin{figure*}
    \centering
    \includegraphics[width=\textwidth]{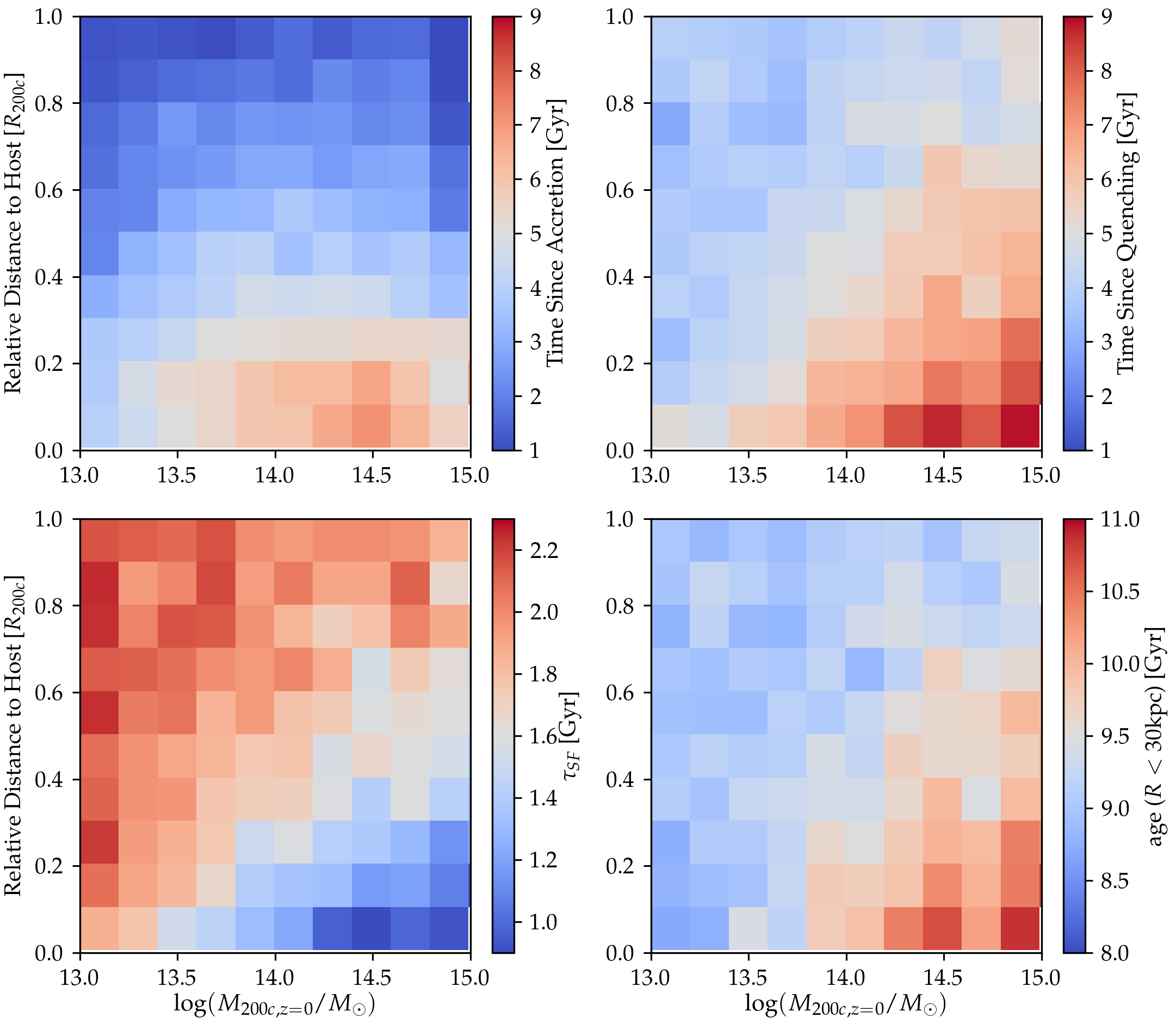}
    \caption[]{ Variation of parameters for TNG300 quiescent satellite galaxies 
    with $10\leq \log{(M_{\star,z=0}/M_{\odot})}<11$ at $z=0$ in the plane of 
    host halo mass at $z=0$ and their relative 3D distance 
    to the primary galaxies.  From top left to bottom right, the parameters are 
    the time since galaxies accreted to the $R_{200c}$ of their final host dark 
    matter halo, the time since galaxies are quenched, the star formation 
    timescales, and the mass-weighted stellar age within $30$~kpc. 
    Colors indicate the mean values in the bins of $0.1R_{200c}$ and $0.2$~dex 
    in cluster-centric distance 
    and halo mass directions, respectively.  The minimum number of galaxies in 
    any bin is 29.} 
\label{fig1}
\end{figure*}

\begin{figure*}[t]
\centering
\includegraphics[width=18cm]{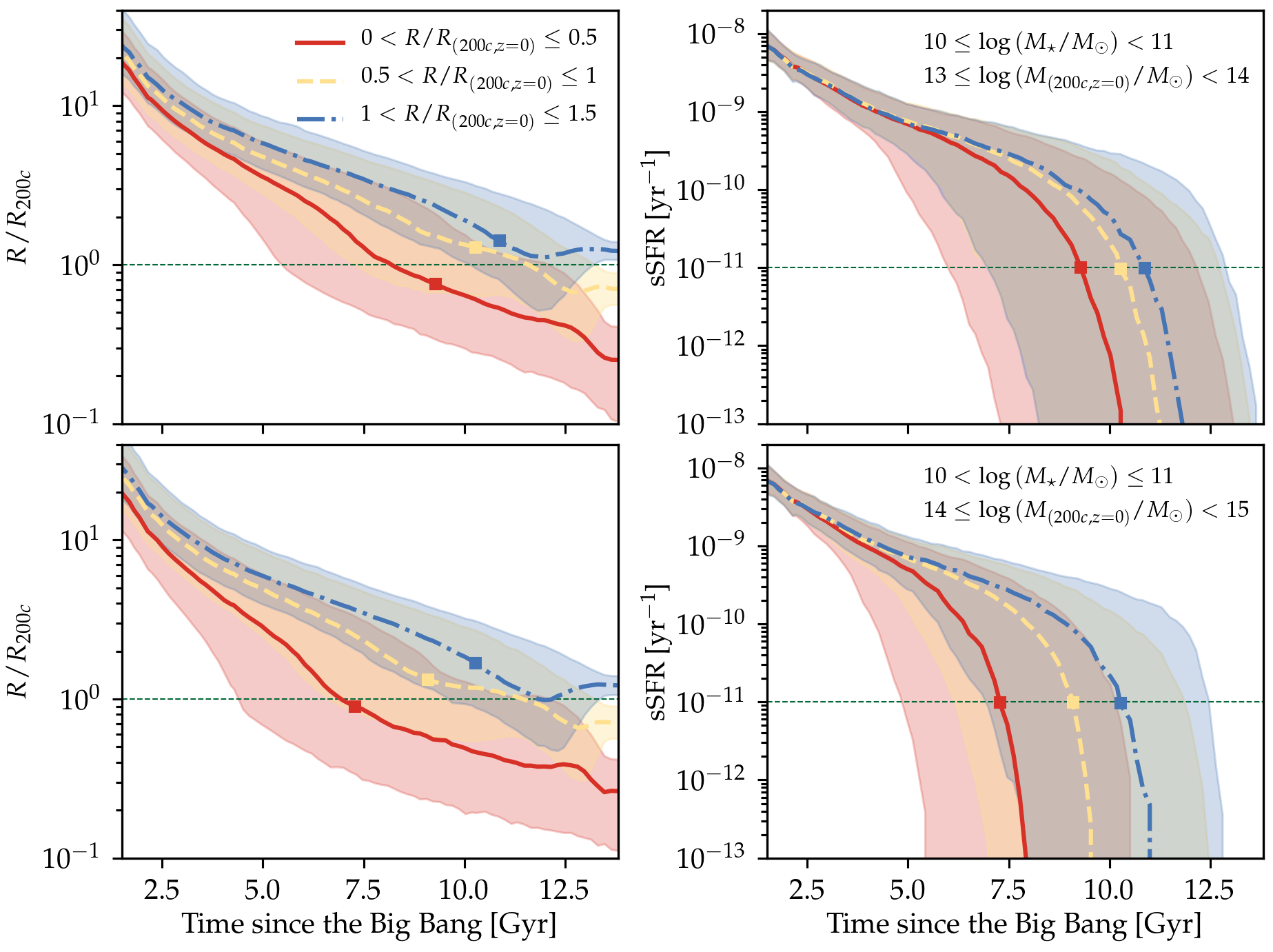}
\caption{
 Evolution of TNG300 quiescent satellite galaxies selected at $z=0$ with 
 $10\leq \log{(M_{\star,z=0}/M_{\odot})}<11$ in two halo mass bins, 
 $13\leq \log(M_{200c,z=0}/M_{\odot})<14$ (top panels) and 
 $14\leq \log(M_{200c,z=0}/M_{\odot})<15$ (bottom panels).  
 Left panels show the evolution of their distances relative to the final 
 primary galaxies at $z=0$, grouped by their relative distances at $z=0$.
 Right panels show the evolution of sSFR in each relative distance group.  
 Lines and shaded regions indicate median and $\pm1\sigma$ values. Squares 
 mark the time that satellite galaxies are quenched.  In both massive clusters 
 and groups, satellite galaxies 
 closer to the halo centers, and in more massive dark matter halos 
 at $z=0$ have earlier accretion histories and quenching histories.  
 }
\label{fig2}
\end{figure*}
\noindent  

\begin{figure*}
\vskip 0.15cm
  \centerline{\psfig{file=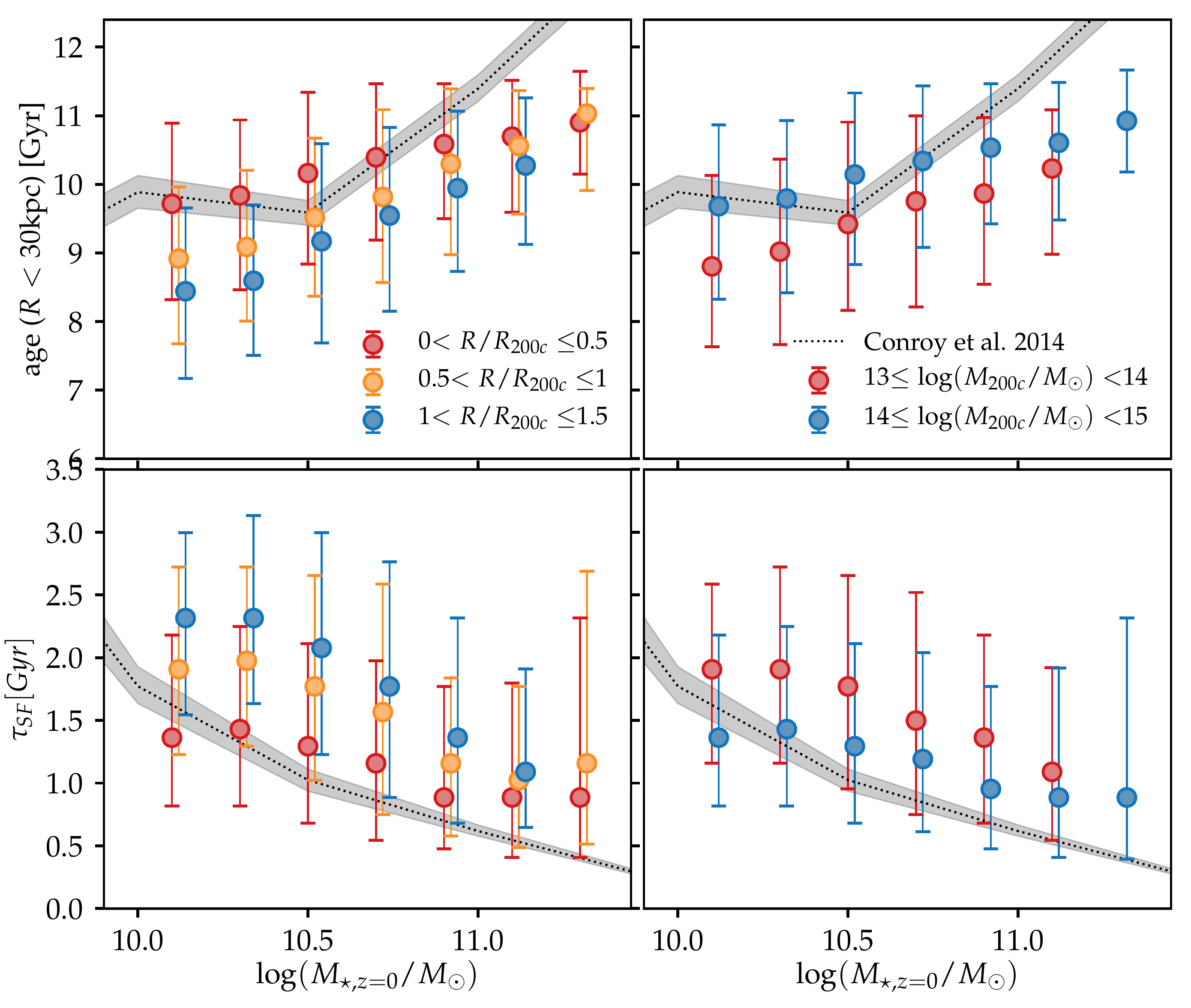, width=18cm}}
  \figcaption[]{
  Scaling relations between stellar mass and stellar ages (top panels), and 
  star formation timescales ($\tau_{SF}$, bottom panels) for satellite galaxies 
  in the TNG300 simulation.  Galaxies are grouped by their relative distance to 
  the primary galaxies in the dark matter halos with 
  $M_{200c}\geq 10^{14}M_{\odot}$at $z=0$ (left panels), or the mass of their 
  final host dark matter halos (right panels).  Colored points indicate the 
  median values and error-bars enclose the $25$th and $75$th percentiles of 
  the simulation results. Observational results are shown in black and gray and 
  are measured from the stacked spectra of SDSS ETGs in all environments 
  \citep{Conroy2014}.  
 }
\label{fig3}
\end{figure*}
\noindent  

\begin{figure*}
\vskip 0.15cm
  \centerline{\psfig{file=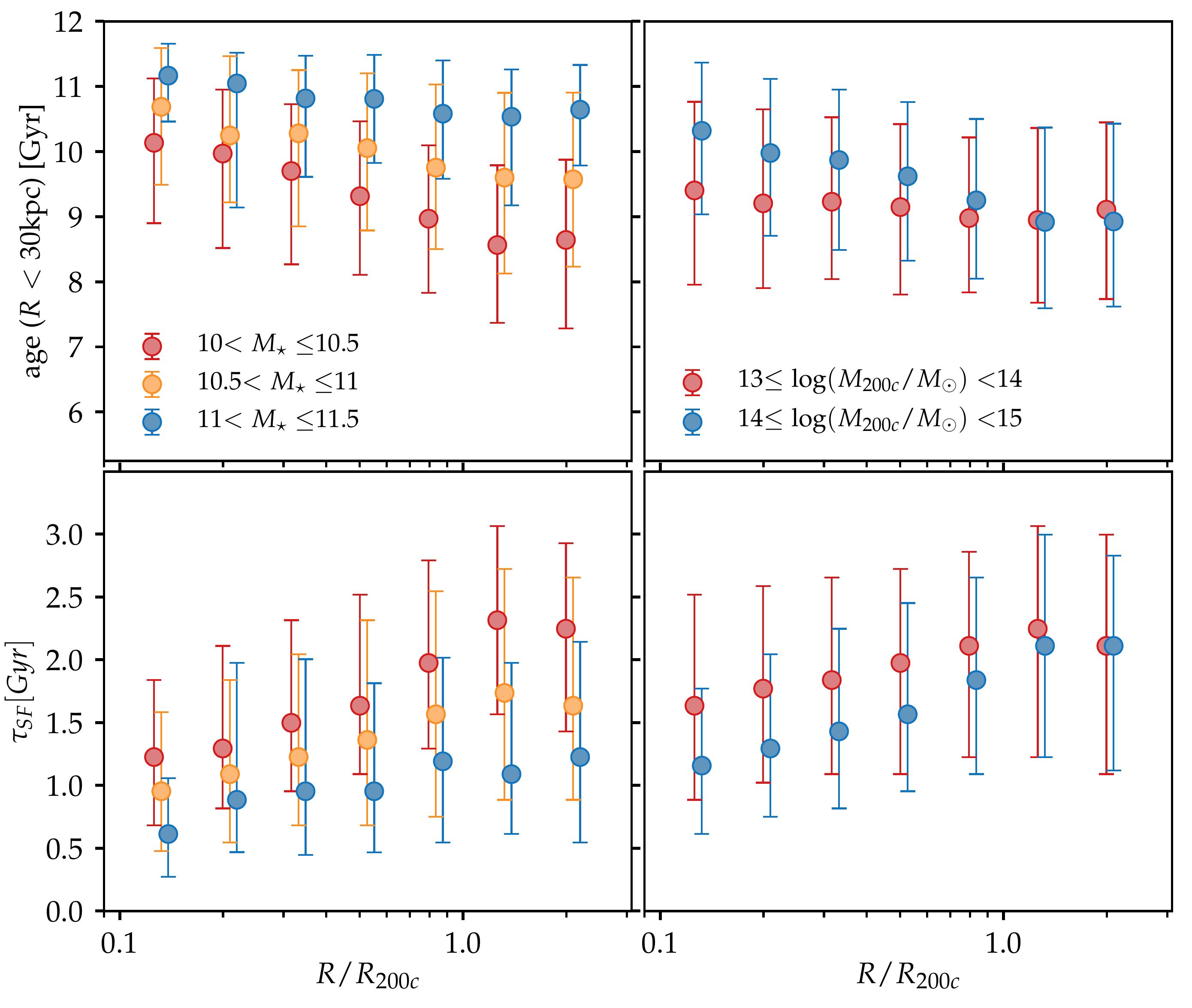, 
  width=18cm}}
  \figcaption[]{ 
  Mass-weighted stellar age and star formation timescales as a function of 
  cluster-centric distances for quiescent satellite galaxies at $z=0$.  
  We focus on galaxies in $\log{(M_{200c,z=0}/M_{\odot})}>14$ in the left 
  panels, and galaxies with $10\leq \log (M_{\star,R<{\rm 30kpc}})<11$ in the 
  right panels.
 }
\label{fig4}
\end{figure*}

\begin{figure*}
\vskip 0.15cm
  \centerline{\psfig{file=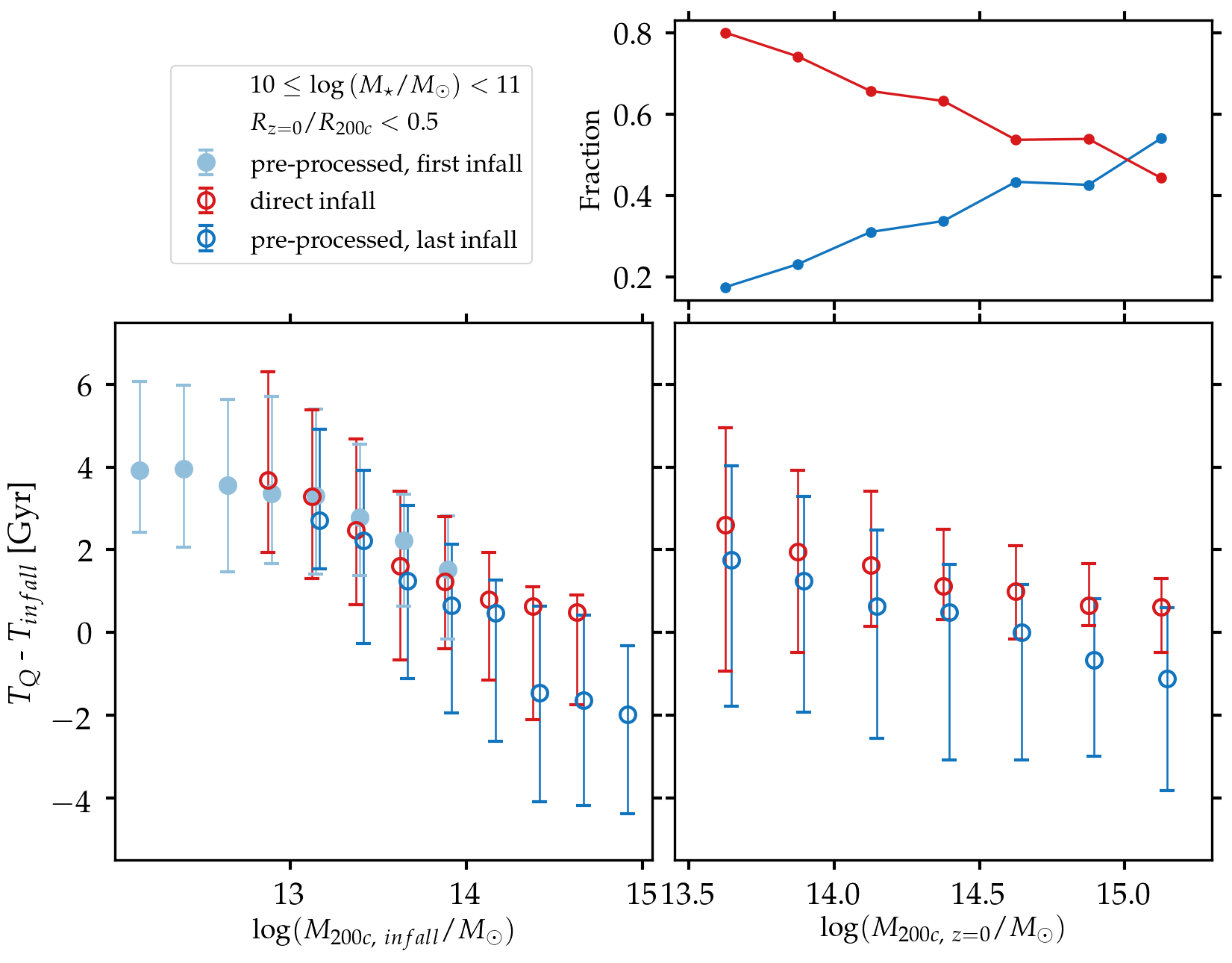, 
  width=18cm}}
  \figcaption[]{Top right: Fraction of satellite galaxies with 
  $10\leq(M_{\star,R<30\rm{kpc},z=0}/M_{\odot})<11$ that have fallen into the 
  final halo directly (red) and that have been group pre-processed (blue). Bottom 
  panels show the time delay between infall and quenching time as a function of 
  the halo mass at the infall time (left), and the halo mass at $z=0$.  
  Data are binned in 0.25~dex in halo mass, and only bins containing more than 
  25 galaxies are shown. Direct infall satellite galaxies experienced $0-3$~Gyr 
  delay between their infall and quenching time.  Galaxies that had experienced 
  group pre-processing in general have a $1-4$~Gyr delay between their first infall 
  and quenching time.  For all galaxies the delay time is shorter in more massive 
  host dark matter halos.
 }
\label{fig5}
\end{figure*}

In a recent paper, \citet{Gu2018b}, we described the ``Coordinated Assembly'' 
picture of massive galaxies using observations of an ongoing brightest cluster 
galaxy (BCG) assembly in Abell~3827. We found that the ETGs in the central 
regions of the galaxy cluster do not follow the [$\alpha/\rm Fe$]-$M_{\star}$ 
and age-$M_{\star}$ trends established by the general ETG sample, but instead 
obey much shallower relations.  Our work indicates that the building blocks of 
massive galaxies in this special environment are different from the overall 
ETG population. As a result, massive central galaxies grow by accreting 
preferentially high [$\alpha/\rm Fe$] and old stellar systems. The flat age 
and [$\alpha/\rm Fe$] radial profiles confirm that the disrupted systems 
should be old and $\alpha$-enhanced.  

This picture highlights the role that environment has on galaxy quenching 
\citep{Peng2010, Peng2012, Wetzel2013, Wetzel2014}:  ETGs in the central 
region of galaxy clusters have ceased their star formation activity earlier 
than their counterparts elsewhere.  Galaxy quenching could be due 
to multiple physical mechanisms, including strangulation 
\citep{Larson1980, Balogh2000}, ram-pressure stripping 
\citep{Gunn1972, Abadi1999}, harassment \citep{Farouki1981, Moore1996}, etc.  
These processes preferentially occur when a galaxy falls into a high 
density environment, but the details about how these mechanisms depend on 
environment, e.g., halo mass or local over-density, is still ambiguous.   

Comparing observations with cosmological hydro-dynamical simulations 
grants us access to a more complete history about how those galaxies form 
in different environments, and allows us to explore the coordinated assembly 
picture in detail.  The IllustrisTNG suites are state-of-the-art 
hydrodynamical simulations that include a large population of massive 
galaxy clusters \citep{Springel2018, Pillepich2018, Pillepich2019, 
Naiman2018, Marinacci2018, Nelson2018a, Nelson2018b, Nelson2019}, and allows 
us to trace the life stories of a statistical sample of massive galaxies 
and their surrounding satellite galaxies in groups and clusters. 
In this paper, we focus on quenched satellite galaxies. The contribution 
from star forming satellite galaxies, although they may be the minority in
clusters, also depends on environment \citep{Donnari2020a}.  
By directly analyzing the star formation, quenching, and assembly histories 
of ETGs in IllustrisTNG, we address the following questions: Does the coordinated 
assembly picture hold in IllustrisTNG?  Are the relevant scaling relations 
consistent with observations?  What is the possible explanation for the 
coordinated assembly picture?

In this Letter, 
we briefly review how we make use of the simulations in Section 2, and 
present the results, including the environmental dependence, and comparison 
with observations in Section 3.  We discuss the importance of group 
pre-processing as well before summarizing our conclusions in Section 4.

\section{Methods}
 
The IllustrisTNG suite is a set of cosmological gravo-magneto-hydrodynamical 
simulations \citep{Springel2018, Pillepich2018, Pillepich2019, Naiman2018, 
Marinacci2018, Nelson2018a, Nelson2018b, Nelson2019}, based on the original 
Illustris simulations \citep{Genel2014, Vogelsberger2014a, Vogelsberger2014b}.  Significant 
improvements in the galaxy formation model provide a more realistic 
prediction of galaxy evolution.  The galaxies are more consistent with 
observations in terms of galaxy sizes and morphological 
types, star formation stellar contents, etc \citep[e.g.][]{Pillepich2018, Nelson2018a, 
Genel2018, Rodriguez-Gomez2018, Diemer2017, Donnari2018}. 
The IllustrisTNG simulations adopt the $\Lambda$~CDM cosmology with 
parameters from \citet{Planck2016}: $h=0.6774$, $\Omega_{m,0}=0.3089$,
$\Omega_{b,0}=0.0486$, $\sigma_8=0.8159$, and $n_s=0.9667$.  
It includes three different cosmological volumes.  In this Letter, we focus 
on the TNG300 run, the run of the IllustrisTNG simulations with the largest box 
with a side length of about 300~Mpc.  The mass resolution of the TNG300 
simulation is $5.9\times 10^7M_{\odot}$ for dark matter and 
$1.1\times 10^7M_{\odot}$ for baryons.  The combination between the simulated 
volume and resolution makes
this run an ideal one for studying the formation and assembly histories of the massive 
galaxies and their satellites in clusters. 

We make use of the halos and subhalos identified by the 
friends-of-friends (FOF) \citep{Davis1985} and {\tt{SUBFIND}} 
\citep{Springel2001, Dolag2009} algorithms, and merger trees 
generated by the {\tt SUBLINK} algorithm \citep{Rodriguez-Gomez2015}. 
Throughout this work,  We focus on groups and clusters with 
$M_{200c}\geq10^{13}M_{\odot}$.  There are 3733 such dark matter halos at $z=0$
up to a cluster of $\sim10^{15}M_{\odot}$, each modeled with dark matter, gas, 
stars, magnetic fields and SMBHs.  The stellar mass of a galaxy refers to the 
total stellar mass of all stellar particles within $30$~kpc.  We employ only
the well-resolved galaxies so the minimum stellar mass in our sample 
is $10^{10}M_{\odot}$.  We focus on quiescent galaxies in this work, and select 
galaxies with a specific star-formation rate (sSFR) less than 
$10^{-11}$yr$^{-1}$ at $z=0$, and only satellite galaxies, excluding
primary galaxies in our sample. The satellite galaxies in this work are all those
within parent FOF groups above the mass threshold, including those located 
outside $R_{200c}$.  There are $615$, $8914$ and $19389$ quiescent satellite 
galaxies with $M_{\star}\geq10^{10}M_{\odot}$ in halos with $M_{200c}$ more than 
$10^{15}M_{\odot}$, $10^{14}M_{\odot}$ and $10^{13}M_{\odot}$, respectively.
We calculate the quenching time, star formation timescale, and the accretion 
time by tracing the main progenitors of galaxies selected at $z=0$ back in 
time. Specifically, we calculate the time since quenching as the time since 
a galaxy's sSFR last fell and remained below $10^{-11}$/yr.  We define 
the time since accretion as the time since a satellite galaxy 
initially fell within $R_{200c}$ of its host dark matter halo.  The star formation 
timescale is calculated using stars formed in-situ, as in 
\citet{Rodriguez-Gomez2016}.  The star formation timescale refers to half 
of the time it takes for the galaxy to form between $16\%$ and $84\%$ of 
its in-situ stars.  The stellar ages in this work refer to the mass-weighted 
stellar ages within $30$~kpc from the galaxy center.  
\\
\\

\section{Results and Discussion}
\subsection{Dependence on Halo Mass and Cluster-centric Distance}

Figure~\ref{fig1} shows the coordinated assembly signal in the TNG300
simulation.  Specifically, the panels show the dependence on environment.  
We focus on quenched satellite galaxies 
in a small stellar mass range at $z=0$, $10\leq \log{(M_{\star,z=0}/M_{\odot})} <11$, 
and show their accretion time (top left), their quenching time (top right), 
the star formation timescale (bottom left) and their mass-weighted 
stellar age (bottom right) in this figure. The color-maps show the mean values 
of these parameters as a function of both the 3D cluster-centric distance at 
$z=0$ in $0.1R_{200c}$ bins, and their host halo mass ($M_{200c}$) in 
$0.2$~dex bins.  The figure reveals a strong dependence on environment for all 
four parameters.  From the top two panels, in general, quenched satellite galaxies 
that live in more massive halos and are located closer to the central galaxies are 
accreted to $R_{200c}$ earlier and quenched earlier, possibly because an early 
accretion means they may get the chance to experience all the quenching mechanisms 
earlier and/or longer than their counterparts.  In fact, as shown in \citet{Rhee2017}, 
satellites at small cluster-centric distances have accreted earlier. Those located 
at $\leq0.2\times R_{200c}$ in massive clusters ($M_{200c,z=0}\geq10^{14}M_{\odot}$) 
at $z=0$ are accreted $\sim 7-8$~Gyr ago and quenched $\sim 6-8$~Gyr ago, while 
those at $\sim R_{200c}$ in low mass dark matter halos 
($10^{13}M_{\odot}\leq M_{200c,z=0}<10^{14}M_{\odot}$) are only 
accreted and quenched $\sim 3-4$~Gyr ago.  Looking at the most massive bins with the 
smallest cluster-centric distance, the trend of time since accretion does not peak 
in the most massive halo mass bin.  There are several reasons for this: 1. the fraction 
of pre-processed galaxies (which will be discussed in Section~3.3) increases 
with increasing mass of the final host halos.  2. On average, satellite galaxies 
that have been pre-processed are accreted into the final host halos at a later time 
than their direct infall counterparts.  3.  For satellite galaxies that have 
been pre-processed, those in a more massive host halos at $z=0$ on average 
have a later accretion time into their final host halos. This 3rd point is not 
statistically strong though, and we need more data to verify it. 

From the color-maps of star formation timescales and stellar ages, galaxies in 
more massive halos and those located closer to the central galaxies are 
formed earlier and cease star formation more abruptly.  This is consistent with 
the coordinated assembly picture revealed by observational stellar population 
analysis: galaxies in the central regions of massive clusters have high 
[$\alpha/\rm Fe$] and old stellar ages.  As a result, the most massive galaxies 
in the centers of the most massive dark matter halos will only get the chance 
to merge with those satellite galaxies that have old stellar ages, earlier 
quenching histories, and short star formation timescales.  The assembled galaxies 
will imprint their stellar populations in the outskirts of massive galaxies, and 
this result can be examined by observational radial profiles of [$\alpha/\rm Fe$] 
and stellar ages.
  
\subsection{Ages and Star Formation Histories}

In Figure~\ref{fig2} we show the evolution of the $z=0$ quenched satellite 
galaxies in different environment.  We focus on galaxies with 
$10\leq \log{(M_{\star,z=0}/M_{\odot})}<11$.
The top two panels show their accretion histories and evolution of sSFR in 
groups 
($10^{13}M_{\odot}< M_{200c,z=0}\leq10^{14}M_{\odot}$), 
and the bottom panels show the evolution in clusters 
($10^{14}M_{\odot}< M_{200c,z=0}\leq10^{15}M_{\odot}$).
Galaxies are grouped by their relative cluster-centric distance at 
$z=0$ into 3 bins: $R\leq0.5\times R_{200c}$ (red), 
$0.5\times R_{200c}<R\leq R_{200c}$ (orange) and 
$R_{200c}<R\leq 1.5\times R_{200c}$ (blue). Squares mark the average time 
that galaxies are quenched.  As shown in Figure~\ref{fig2},  in both 
group and cluster environments, galaxies that are located closer to the 
primary galaxies have earlier accretion histories.  Their sSFR falls 
faster and they are quenched earlier on average.  Comparing galaxies 
with similar relative cluster-centric distances in group and cluster environments 
reveals that those in clusters have earlier 
accretion and quenching histories. Some galaxies are quenched even before 
they enter the $R_{200c}$ of their final host dark matter halos, 
indicating that their gas has been stripped before entering 
their final host dark matter halos or pre-processed in smaller groups.


In Figure~\ref{fig3}, we compare the $\tau_{SF}$-$M_{\star}$ and stellar 
age-$M_{\star}$ scaling relations of the TNG300 quiescent galaxies 
with observed quiescent galaxies.  Black dotted lines and the 
corresponding gray regions show observational results of stellar age and 
star formation timescale ($50$th, $16$th and $84$th percentiles) 
based on stacked SDSS early-type galaxies (ETGs) that are binned in stellar mass 
\citep{Conroy2014}.  The stacked spectra are fit with the upgraded response 
functions.  The star formation timescales are estimated using 
[$\alpha/\rm Fe$] of these stacked ETGs, and the conversion between 
[$\alpha/\rm Fe$] and star formation timescale based on a simple chemical evolution 
model in \citealt{Thomas1999, Thomas2005}:
  $[\alpha/\rm Fe] \approx {\frac{1}{5}} - {\frac{1}{6}} \log{(\Delta t/{\rm Gyr})} $. 
We examine how these scaling relations depend on cluster-centric distance 
(left panes, focusing on $\log{(M_{200c}/M_{\odot})}>14$) and halo mass 
(right panels, focusing on $R\leq 0.5\times R_{200c}$) using TNG300 galaxies.  
The trends are calculated in 0.2~dex bins of $M_{\star,R<{\rm 30kpc}}$.  Only bins 
including more than 25 galaxies are shown in the figure.  The overall 
trends followed by TNG300 quiescent galaxies are in good agreement with 
observations.  In both previous observations and this work, more massive 
galaxies are older and have short star formation timescales.  
More importantly, the TNG300 simulations predict strong variation in 
different environments: galaxies in the central regions and in more massive halos 
have shallower $\tau_{SF}$-$M_{\star}$ and stellar age-$M_{\star}$ 
relations, indicating that in special environments, such as 
the central regions of galaxy clusters, even the relatively low 
mass galaxies have old stellar ages, and short star formation 
timescales in the past.  This is consistent with the coordinated 
assembly picture.  

Figure~\ref{fig4} shows the stellar ages and star formation timescales 
of quenched satellite galaxies as a function of relative cluster-centric 
distance in different environments.  We focus on galaxies in 
$\log{(M_{200c,z=0}/M_{\odot})}>14$ in the left panels, and 
galaxies with $10\leq \log{(M_{\star}/M_{\odot})}<11$ 
in the right panels. The general trends are:  
satellite galaxies located further away from the primary galaxies are 
younger, and have longer star formation timescales in the past. 
From the left panels, the dependence on cluster-centric distance is 
more robust for low mass galaxies.  The left panels also 
predict that for galaxies in clusters with similar mass, the relation 
between stellar age and/or [$\alpha/\rm Fe$] and cluster-centric distance 
may have larger scatter towards the outskirts of galaxy clusters.
From the right panels, galaxies in more massive dark matter halos are 
older and on average formed faster in the past, but this 
effect is only strong within $R_{200c}$.  
  
\subsection{Group Pre-processing}

Previous work has highlighted the importance of group pre-processing on 
galaxy quenching histories \citep[e.g.][]{Fujita2004,Balogh2010,Wetzel2013,Bahe2019}.
In Figure~\ref{fig5}, we look into the dependence on halo mass in detail, and examine whether group pre-processing plays a role in the 
coordinated assembly picture.  
We focus on satellite galaxies with $10\leq \log{(M_{\star,z=0}/M_{\odot})}<11$ 
that are located relatively close to the cluster centers 
($\log{(M_{200c,z=0}/M_{\odot})}>14, R<0.5\times R_{200c}$).  We identify satellite 
galaxies that fall into their $z=0$ halos directly as central galaxies, 
and include them in the ``direct infall'' groups in Figure~\ref{fig5}. 
The first time they are accreted into $R_{200c}$ of their final host dark 
matter halos are marked as their infall times (red).
Next, we identify those that were accreted onto their $z=0$ halos as satellites, and look into 
the most recent time they fell into any dark matter halos with 
$\log{(M_{200c,z=0}/M_{\odot})}>12$ as central galaxies.  We include 
these galaxies as ``pre-processed'' galaxies, and mark the time they are 
accreted as central galaxies as their first infall time (light blue), 
and the time they are accreted into their final host dark matter halos as the 
last infall time (blue).
We note that one caveat comes from the uncertainty in 
merger scenarios where position-space halo finders may not successfully trace 
the separation of all subhalos.

The top right panel of Figure~\ref{fig5} shows the fraction of direct infall 
and group pre-processed galaxies as a function of their host dark matter halo 
mass at $z=0$.  In halos with $\log{(M_{200c,z=0}/M_{\odot})}\sim 13.5$, 
$80\%$ galaxies are accreted as central galaxies, and only $<20\%$ have 
been pre-processed by a smaller halos.  However, group pre-processing 
plays an increasingly important role in more massive galaxies.  
In the massive clusters with $\log{(M_{200c,z=0}/M_{\odot})}\approx 15$, 
more than half the quenched galaxies have experienced group pre-processing.  

In the bottom panels, we plot the delay between the quenching and infall times 
as a function of the halo mass at the infall time (left) and the halo mass 
at $z=0$ (right).  As shown in the bottom left panels, the time delay is shorter 
when galaxies are accreted into more massive dark matter halos.  More importantly, 
the slopes of the relations between the time delay and the first infall 
halo mass are similar in both the ``direct infall'' 
and ``pre-processed'' cases, indicating that the quenching processes have 
already started since the first infall.  If we only compare the time delay 
since the last infall, galaxies that have experienced group pre-processing 
 have much shorter delay times 
than the direct infall galaxies, some of which are quenched even 
before they are accreted.  This panel shows that group pre-processing 
is crucial in satellite quenching.  The bottom right panel shows the 
time delay since their final accretion as a function of $z=0$ halo mass.  
Galaxies that have been pre-processed have shorter time delays ($\approx -1-2~$Gyr) 
at all halo masses, while the direct infall galaxies are quenched $0-3$~Gyr 
after they are accreted.  In both cases galaxies are quenched faster in 
more massive groups and clusters. See \citet{Donnari2020a} 
for additional in-depth characterization of pre-processing 
with the IllustrisTNG simulations.


\section{Summary}
In this Letter, we use TNG300 galaxies to explore the coordinated assembly 
picture \citep{Gu2018b}, namely the idea that the ETGs in the central regions of galaxy 
clusters do not obey the [$\alpha/\rm Fe$]-$M_{\star}$ and age-$M_{\star}$ 
trends established by the general ETG sample, but instead follow much 
shallower relations.  We try to understand whether the simulations show 
coordinated assembly and, if they do, what could be the underlying physical 
cause.  We focus on quiescent satellite galaxies with 
$\log{(M_{\star,z=0}/M_{\odot})}\geq10$ in massive halos 
($M_{200c,z=0}\geq10^{13}M_{\odot}$) from the TNG300 run.
Our conclusions are summarized as follows:

\begin{itemize}
\item At $z=0$, quenched satellite galaxies in the central regions of massive 
groups and clusters are accreted earlier, quenched earlier, have shorter 
star formation timescales and older stellar ages compared to their counterparts 
in other environments.  This is consistent with observational results 
that galaxies in the central regions of massive clusters are old and 
$\alpha$-enhanced.

\item The star formation timescale ($\tau_{SF}$)-$M_{\star}$ 
and stellar age-$M_{\star}$ scaling relations 
followed by the TNG300 galaxies are in good agreement with observations.  
The slopes of these relations are much shallower in the central regions of massive 
dark matter halos, indicating that in this special environment even the 
low-mass ETGs have old stellar ages, and short star formation timescales.

\item Both stellar ages and star formation timescales depend on cluster-centric 
distance.  Galaxies in more massive dark matter halos are older and formed faster 
on average within $R_{200c}$.  The dependence on cluster-centric distance 
is stronger for low mass galaxies.

\item Group pre-processing is a crucial aspect in galaxy quenching, and plays 
an increasingly important role in more massive halos.  Galaxies accreted into more 
massive groups and clusters are quenched faster, and the time delay is primarily 
determined by the halo mass of the first accretion.

\end{itemize}

To test the prediction of the environmental dependence in this work with observations, 
both halo mass and cluster-centric distance should be 
carefully measured from the observational side
so that a robust comparison can be made. 
During the assembly of massive galaxies, the building blocks will imprint 
their stellar population signatures in the outskirts of these galaxies.  
Investigating the radial profiles of stellar population properties such 
as age and $\alpha$-abundance in different environments, in both observations 
and simulations, will test the coordinated assembly picture.

\acknowledgments
This work made use of the {\tt COLOSSUS} package \citep{Diemer2018}.
M.G. acknowledges support from the National Science Foundation Graduate 
Research Fellowship and the Henry Norris Russell Postdoctoral Fellowhip.  
C.C. acknowledges support from NASA grant 
NNX15AK14G, NSF grant AST-1313280, and the Packard Foundation. 
FM acknowledges support through the Program 
``Rita Levi Montalcini'' of the Italian MIUR.
The computations in this paper were run on the 
Odyssey cluster supported by the FAS Division of Science, Research 
Computing Group at Harvard University.   
The IllustrisTNG simulations used in this work have been run on the 
HazelHen Cray XC40-system at the High Performance Computing Center 
Stuttgart as part of project GCS-ILLU of the Gauss Centres for 
Supercomputing (GCS).


\end{document}